\documentclass[sigconf]{acmart}
\def\BibTeX{{\rm B\kern-.05em{\sc i\kern-.025em b}\kern-.08emT\kern-.1667em\lower.7ex\hbox{E}\kern-.125emX}}

\usepackage{nicefrac}
\usepackage{siunitx}
\usepackage{array,framed}
\usepackage{booktabs}
\usepackage{
  color,
  float,
  epsfig,
  wrapfig,
  graphics,
  graphicx,
  subcaption
}
\usepackage{textcomp}
\usepackage{setspace}
\usepackage{latexsym,fancyhdr,url}
\usepackage{enumerate}
\usepackage{algorithm2e}
\usepackage{algpseudocode}
\usepackage{graphics}
\usepackage{xparse} 
\usepackage{xspace}
\usepackage{multirow}
\usepackage{csvsimple}
\usepackage{balance}

\usepackage{color}
\usepackage{colortbl}
\usepackage{multirow}

\usepackage{
  tikz,
  pgfplots,
  pgfplotstable
}
\usepackage{hyperref}

\usetikzlibrary{
  shapes.geometric,
  arrows,
  external,
  pgfplots.groupplots,
  matrix
}

\setcopyright{none}
\renewcommand\footnotetextcopyrightpermission[1]{}

\pgfplotsset{compat=1.9}

\usepackage{mathtools,}

\DeclareMathAlphabet{\mathcal}{OMS}{cmsy}{m}{n}

\DeclareGraphicsExtensions{%
    .png,.PNG,%
    .pdf,.PDF,%
    .jpg,.mps,.jpeg,.jbig2,.jb2,.JPG,.JPEG,.JBIG2,.JB2}

\usepackage{xparse}
\newcommand{\bnm}{\begin{newmath}}
\newcommand{\enm}{\end{newmath}}

\newcommand{\bea}{\begin{eqnarray*}}%
\newcommand{\eea}{\end{eqnarray*}}%

\newcommand{\bne}{\begin{newequation}}
\newcommand{\ene}{\end{newequation}}

\newcommand{\bal}{\begin{newalign}}
\newcommand{\eal}{\end{newalign}}

\newenvironment{newalign}{\begin{align}%
\setlength{\abovedisplayskip}{4pt}%
\setlength{\belowdisplayskip}{4pt}%
\setlength{\abovedisplayshortskip}{6pt}%
\setlength{\belowdisplayshortskip}{6pt} }{\end{align}}

\newenvironment{newmath}{\begin{displaymath}%
\setlength{\abovedisplayskip}{4pt}%
\setlength{\belowdisplayskip}{4pt}%
\setlength{\abovedisplayshortskip}{6pt}%
\setlength{\belowdisplayshortskip}{6pt} }{\end{displaymath}}

\newenvironment{newequation}{\begin{equation}%
\setlength{\abovedisplayskip}{4pt}%
\setlength{\belowdisplayskip}{4pt}%
\setlength{\abovedisplayshortskip}{6pt}%
\setlength{\belowdisplayshortskip}{6pt} }{\end{equation}}

\newcounter{ctr}

\newcounter{mytable}
\def\mytable{\begin{centering}\refstepcounter{mytable}}
\def\endmytable{\end{centering}}

\newcounter{myfig}
\def\myfig{\begin{centering}\refstepcounter{myfig}}
\def\endmyfig{\end{centering}}

\newlength{\saveparindent}
\newlength{\saveparskip}
\newcommand{\E}{{\rm I\kern-.3em E}}

\renewcommand{\eqref}[1]{\mbox{Equation~(\ref{#1})}}

\def \part {part}

\renewcommand{\paragraph}[1]{\vspace*{6pt}\noindent\textbf{#1}\;}

\def \blackslug{\hbox{\hskip 1pt \vrule width 4pt height 8pt
    depth 1.5pt \hskip 1pt}}
\def \qed{\quad\blackslug\lower 8.5pt\null\par}

\newcounter{mynote}[section]

\newcommand\ignore[1]{}

\newcounter{rcnote}[section]

\newcounter{mrnote}[section]

\newcounter{fknote}[section]

\newcounter{anote}[section]

\DeclareMathSymbol{\mlq}{\mathord}{operators}{``}
\DeclareMathSymbol{\mrq}{\mathord}{operators}{`'}

\newcommand{\rhf}[2]{R_{f, \gamma}}

\DeclareDocumentCommand{\edist}{o o}{
  \ensuremath{
    \IfNoValueTF{#1}{{d}}{{\sf d}(#1,#2)}
  }
}

\newcommand{\olrk}[1]{\ifx\nursymbol#1\else\!\!\mskip4.5mu plus 0.5mu\left(\mskip0.5mu plus0.5mu #1\mskip1.5mu plus0.5mu \right)\fi}

\NewDocumentCommand{\indseq}{ O{1} O{r} }{{#1}\ldots {#2}}

\begin{document}
\fancyhead{}
\def\thetitle{Measuring Miner Decentralization in \\ Proof-of-Work Blockchains}
\title{\thetitle}

\author{Sishan Long}
\affiliation{%
    \institution{Cornell Tech; Cornell University; IC3}
}
\email{sl3275@cornell.edu}

\author{Soumya Basu}
\affiliation{%
    \institution{Cornell University; IC3}
}
\email{soumya@cs.cornell.edu}

\author{Emin G{\"u}n Sirer}
\affiliation{%
  \institution{Ava Labs}
}
\email{egs@systems.cs.cornell.edu}

%

\date{}

\begin{abstract}
Proof of work cryptocurrencies began with the promise of a more egalitarian future with a decentralized monetary system
    with no powerful entities in charge.
While this vision is far from realized, these cryptocurrencies are still touted to be much more
    decentralized than traditional centralized systems.
While it is well understood that cryptocurrencies are centralized, it is still unclear what the underlying causes
    are.
This work aims to address this gap and examines some of the forces behind mining centralization.

The internals of cryptocurrency mining are very opaque and difficult to study since it traditionally requires
    forming relationships with miners, who are typically reticent to share internal information about their competitive
    advantages.
This work takes a different approach by combining large scale statistical techniques with publicly available blockchain
    data in order to answer previously intractable questions.
The crux of our analysis technique is based on the simple observation that some miners can utilize their hashpower more
    efficiently due to their position in the network.
By teasing out that effect, we de-bias the mining power distribution to get a more accurate estimate.
Using that de-biased mining power distribution, we can answer questions about the network position of miners in
    each cryptocurrency network.
Finally, during the course of this study, we observed some unusual mining behaviors which we highlight.

\end{abstract}

\maketitle

\section{Introduction}

\newcommand{\tabincell}[2]{\begin{tabular}{@{}#1@{}}#2\end{tabular}}
\begin{table*}[pthb]
  \begin{center}
  \begin{tabular}{|c|c|c|c|c|c|}
    \hline
      Coin & \tabincell{c}{Max Block Size (bytes)} & \tabincell{c}{Avg Block Size (bytes)} & \tabincell{c}{Avg Block Interval (sec)}& \tabincell{c}{Avg Goodput\\(bytes/sec)}  & \tabincell{c}{Max Goodput\\(bytes/sec)} \\
    \hline
    BTC & 2259447 & 812292 & 570 & 1423 & 1604075\\
    BCH & 23158974 & 190339 & 601 & 322 & 329284\\
    BSV & 255996865 & 820084 & 603 & 1378 & 4266445\\
    ETH & 135880 & 22682 & 14 & 1554 & 91994\\
    ZEC & 1998373 & 17178 & 114 & 115 & 73787\\
    LTC & 995122 & 26818 & 151 & 188 & 15558\\
    \hline
\end{tabular}
 \caption{These are summary statistics of the blocks collected from each blockchain.
      The goodput is defined as the number of bytes added to the blockchain per second.}
  \label{tab:blockchainlimit}
  \end{center}
\end{table*}

Proof of work cryptocurrencies are touted to be a trustless, censorship-resistant form of money.
The key value proposition is that there is no individual or small group of entities that have control over the entire
    currency.
In particular, there must necessarily be many, non-colluding miners that are processing transactions in order to prevent
    any small group of them to enact policies that may break censorship resistance.
If there are only a few miners, then these systems would be no different from an inefficient version of a centralized
    database.

The degree of mining centralization has been previously studied~\cite{adem18,meiklejohn,giniblogpost}, and some of its
    underlying causes are understood.
For example, individual miners form mining pools and share resources to reduce the variance of their block rewards.
Another source of mining centralization is the efforts miners spend to be well positioned in the mining network,
    including making manual connections to many nodes, multiple points of presence, and even peering relationships between
    large miners to relay newly mined blocks more quickly between them.
While these behaviors are well-known, the degree to which these behaviors affect mining centralization is not well
    understood.

Understanding the effect of these behaviors requires making inferences about the mining network.
However, direct measurements on mining networks are extremely difficult to conduct since miners are privacy oriented.
Indirect measurements to infer network properties is also often difficult and expensive~\cite{Miller2015DiscoveringB}.
This paper takes a different approach using large scale statistical techniques on readily available blockchain data to 
    extract insights about the mining network.
Blockchain data has the benefits of direct measurement, as the fields of a block are filled by the miner of that block,
    without the difficulty of deploying a measurement probe and convincing miners to run it.

The crux of our approach rests on the following simple, but powerful, observation.
Per the Nakamoto consensus rules, a miner will transmit their newly mined block to all other miners in the network.
However, the transmission of their freshly mined block takes time, dependent on their position in the network.
Thus, the miner of a newly mined blocks gains an advantage over all other miners since it receives its own block first,
    an advantage we call the \emph{previous block advantage}.
Thus, by looking at the blocks that were mined immediately after a particular miner, we can infer how well connected
     that miner is to all other miners in the network.
A particular advantage to this approach is that it relies on publicly available blockchain data which makes it feasible
    for us to study and compare six different cryptocurrency networks, the largest such study to date.

We proceed with our measurement study in three distinct phases.
The first phase is the preliminary measurements and inferences.
We confirm that the hypothesized previous block advantage indeed exists in the publicly available blockchain data.
Once confirmed, we note that the mining power distribution is biased due to the previous block advantage, and we present
    a more unbiased mining power estimate.
An accurate mining power distribution is useful for the community at large to detect when a miner is amassing a large
    amount of hashpower and for miners to see how much they are underutilizing their hashpower.

Then, our study moves on to answering detailed questions about the mining network and its block propagation characteristics.
Understanding block propagation is key to developers so they can spend their energy optimizing the correct bottlenecks
    and make informed decisions about key protocol parameters such as the block size.
Such decisions are very difficult as some cryptocurrencies, such as Bitcoin (BTC) and Ethereum (ETH), have much faster 
    networks than others that have larger block size limits, such as Bitcoin Cash (BCH) or Bitcoin SV (BSV).
The block size debate is extremely contentious and is one of the reasons behind multiple chain splits.
Previous studies~\cite{adem18,onscalingdecentralized} have proposed setting the blocksize by measuring the bandwidth
    of the entire peer to peer network.

Our work takes an alternate approach to the block size question by looking at the previous block advantage.
Larger blocks propagate through the network more slowly, so the position of a miner in the network confers more of an
    advantage compared to smaller blocks.
However, depending on the network capacity, the threshold for how large a block has to be to confer such an advantage
    changes.
This threshold is called the \emph{safe envelope} of a cryptocurrency, which is formally defined as the maximum block
    size that gives the same previous block advantage as an empty block.
We estimate this threshold for all cryptocurrencies and present a comparative analysis to show the effect of Ethereum
    network upgrades in December 2019, which have helped decrease the previous block advantage.

Finally, we examine some unusual miner behaviors in the mining network, namely the formation of mining cartels and
    active chain switching.
We search and find examples of miner cartelization in ZCash and Litecoin, where mining pools are only nominally separated 
    into separate entities and functionally act as a single larger mining pool.
Cartels are dangerous as they provide the illusion of decentralization rather than true decentralization.
We also examine one case of active chain switching, a behavior by miners to switch between chains frequently in order to
    generate the maximum returns on their mining power.
Due to space constraints, we summarize the high level takeaways for the unusual miner behaviors and defer detailed
    discussion to the appendix.

Our work illustrates how much data there is contained in the blockchain itself and how we can leverage that data to
    answer otherwise intractable questions about the mining network.
Such techniques have a significant advantage over conventional measurement techniques since they do not require
    extensive, custom-built measurement probes which allow for larger-scale studies.
Through these techniques, this work was able to extract a more accurate estimate of the mining power distribution, 
    provide guidance to developers on how to measure the effects of an upgrade and set important protocol parameters, 
    and find examples of unusual mining behavior.

\section{Background}

While cryptocurrencies today have drastically different designs~\cite{algorand,thundercore,ava}, our study focuses on
  the subset that uses proof-of-work based Nakamoto consensus~\cite{nakamoto2008bitcoin,garay2015bitcoin,ethereum,zerocoin}.
In this section, we cover some of the more pertinent aspects of these cryptocurrencies to our study.

\subsection{Proof of Work Cryptocurrencies}

Cryptocurrencies sequence transactions so that all honest participants can agree on the history of transactions.
This agreement is reached in the presence of Byzantine actors in the network, as long as the Byzantine actors are sufficiently weak.
In particular, Nakamoto consensus assumes that Byzantine actors are limited to using at most a quarter of the total
    compute resources in the network~\cite{selfishmining}.

Transactions are sequenced in batches, called \emph{blocks}.
These blocks are ordered by \emph{miners} who append each batch to the tail of the \emph{blockchain} so that all blocks
    are consistently ordered.
To append a block, miners must first decide on its contents and then solve a crypto puzzle in order to prove that they
    have spent a large amount of computing resources.
Our study relies on two pieces of information that are attached to each block: the identity of the miner and the
    timestamp.
Miner identities are self-reported by the miners and are not enforced by the protocol.
Miners gain influence the more blocks they mine, so, like prior work~\cite{adem18,Nayak2015StubbornMG,wang2015exploring},
    we assume that miner identity information is accurate.

Occasionally, miners attempt to insert a block into the same location on the blockchain.
Such events, called \emph{forks}, occur when one miner fails to hear of a recent block and has mined a block on top of 
    a stale blockchain.
Some forks are inevitable in Nakamoto consensus and most cryptocurrencies resolve forks through the longest chain rule,
    where the longest chain of blocks is considered valid.
Most cryptocurrencies drop forked blocks as they are no longer part of the agreed-upon history of events.
Ethereum, on the other hand, saves these forked blocks and rewards miners of forked blocks using the GHOST~\cite{ghost}
    protocol, though the reward for a forked block is significantly smaller.
How often forks occur depends on how long blocks take to propagate and the block interval, which is how often blocks are
    mined.
Table~\ref{tab:blockchainlimit} presents summary statistics on how large each system's blocks are and the corresponding
    effective throughput of the chain.

\begin{table*}[tph]
    \centering
    \begin{tabular}{|c|c|c|c|c|c|}
      \hline
    Coin & ~Start Date~ & ~End Date~ & ~Start Height~ & ~Total Blocks~ & Blocks Dropped\\
      \hline
    BTC & 2018-06-01 & 2018-09-30 &  525367 & 18468 & 3599\\
    BCH & 2018-06-01 & 2018-09-30 & 532682 & 17523 & 4092\\
    BSV & 2019-06-01 & 2019-09-30 &  584862 & 17481 & 51\\
    ETH & 2018-08-01 & 2018-09-15 &  6065980 & 273106 & 64560\\
    ZEC & 2018-08-01 & 2018-09-15 &  368197 & 26399 & 1444\\
    LTC & 2018-08-01 & 2018-09-15 &  1466383 & 26305 & 14282\\
    ETH (new) & 2019-10-25 & 2019-11-07 & 8806000 & 86979 & 10069\\
      \hline
\end{tabular}
    \caption{This shows the dates of the snapshots analyzed in this work 
    as well as the precise block height the dataset
    started at, how many blocks were collected, and how many were dropped to sanitize the data. ETH (new) is a snapshot of the Ethereum blockchain after a new
    network client was deployed.}
  \label{tab:data}
\end{table*}

\subsection{Miner Behaviors}
Miners are rewarded when their blocks are incorporated into the blockchain.
As a result, miners innovate to find novel ways to transmit blocks to and from all other miners.
Miners, especially large miners, often connect directly to each other to transmit block information and
    optimistically start mining on top of a newly received block without fully validating its contents.
These relationships are not necessarily symmetric as well, as a small miner may be willing to optimistically mine a
    block from a larger miner while the reverse may not be true.
These relationships are typically forged directly between miners and not openly accessible, which makes
    inferring properties about the structure of the mining network extremely difficult.
This makes answering questions about mining centralization hard.

Miner behavior also becomes more complex when considering the modern cryptocurrency ecosystem.
When Bitcoin~\cite{nakamoto2008bitcoin} was first released, it was the only proof of work cryptocurrency in existence, and miners
    consisted of home users mining on their laptops.
Modern miners use specialized hardware to mine most large cryptocurrencies and have a choice in which cryptocurrency to
    mine.
Some coins, such as Bitcoin (BTC), Bitcoin Cash (BCH), and Bitcoin SV (BSV), share the same crypto puzzle so specialized
    hardware built for one coin can be redirected fairly easily to another coin.
Thus, miners may choose to switch between coins every so often depending on how profitable it is to mine each one.

\section{Measurement Data}
Our analysis relies on readily available blockchain data and uses large scale statistical analysis to 
    extract insights.
In a given cryptocurrency, we extract the following data from each block: (1) the \emph{block height}, which 
    is determined 
    by counting the number of blocks preceding it in the blockchain, (2) the \emph{block size}, which is the number of
    bytes in a block, (3) the \emph{block timestamp}, which is the time that
    the block was mined, and (4) the identity of the miner who mined this block.
This information is readily available on any block explorer for a cryptocurrency.
We collected data on Bitcoin (BTC), Bitcoin Cash (BCH), Bitcoin SV (BSV), Ethereum (ETH) and Litecoin (LTC) from
    BlockChair~\cite{https://blockchair.com/}.
ZCash (ZEC) data was obtained from the ZChain~\cite{https://api.zcha.in/} block explorer.
Summary statistics about the block data used in our study are presented in Table~\ref{tab:data}.

Our target measurement period is August 2018 for most coins and August 2019 for BSV, as BSV did not exist in 2018.
As our statistical techniques require larger amounts of data, we extend our measurement period slightly before and
    after that time.
We choose the smallest time period as we assume that the mining hashpower is relatively stable so that the differences
    we observe can be attributed to the network.

Two of the fields we extract from the blockchain have potential inaccuracies: the miner identities and the block
    timestamps, which are both set by the miner of the block.
Similar to prior work, we believe the miner identities are accurate as large miners are able to negotiate preferable 
    peering relationships and otherwise influence the network, so it in their best interest to claim their own
    blocks~\cite{adem18,Nayak2015StubbornMG,wang2015exploring}.
We validated the block timestamp data by checking whether the timestamp field matched the time that the block
    was received by the block explorer and ensuring that they were similar.
As an additional validation step, we ran a full node on some networks such as Bitcoin (BTC) and Ethereum (ETH)
   and verified that the block was received at a time similar to its timestamp.

To sanitize our data, we drop certain types of blocks.
Smaller miners, which we define as having mined less than $1\%$ of the total number of blocks in our sample, need to be
    dropped as our analysis requires a large amount of data and small miners just have not produced very many blocks.
Additionally, some blocks are unclaimed and they also get dropped.
These blocks are typically mined by smaller miners, so we do not expect that ignoring unclaimed blocks will affect the
    soundness of our analysis.
We use all blocks in one analysis where we are comparing the relative decentralization of all cryptocurrencies in
   Section~\ref{sec:miningpower}.
The number of blocks dropped in each cryptocurrency during sanitization is noted in Table~\ref{tab:data}.
Note that dropping blocks does not present an error in our measurement, as our inferences are on the miners that were
    included in our sample.

\section{The Previous Block Advantage}
When a miner mines a block, it receives the most recent copy of the blockchain first and obtains an advantage in the
    race to mine the next block, a phenomenon we call the \emph{previous block advantage}.
The previous block advantage encourages mining centralization as miners may choose to coalesce into larger miners or 
    form peering relationships in order to gain a more advantageous position in the network.
Thus, it is crucial to understand and quantify the mechanics of the previous block advantage.

To understand the previous block advantage and its effect on decentralization, we first need to quantify what we mean
    by decentralization.
We use a metric based on the Shannon entropy~\cite{shannon} to compare how decentralized the mining power is between
    different cryptocurrencies.
Then, we verify whether or not the previous block advantage exists in our dataset and introduce a metric to quantify
    how prevalent it is across cryptocurrencies.
Finally, we eliminate the effect of the previous block advantage on the mining power distribution in order to get
    a more accurate estimate of the miner's true hashpower.
This allows us to examine the state of the mining network more carefully in subsequent sections.

\subsection{Mining Power Distribution}
\label{sec:miningpower}

The \emph{mining power distribution} of each coin is the proportion of hashpower owned by each miner on that coin.
A miner with enough hashpower can gain control of the cryptocurrency, censor transactions, and ruin the value proposition
    of the cryptocurrency.
To measure decentralization, we compare the randomness present in the mining power distribution with uniform
    distribution with $n$ miners.
The larger $n$ is, the more decentralized the mining power distribution is.
To compute $n$, we take two to the power of the Shannon entropy~\cite{shannon} in a manner similar to how the effective
    anonymity set size is computed in anonymous communication systems~\cite{danezis}. 

\begin{table}
\center
  \begin{tabular}{|c|c|}
   \hline
    Coin & \tabincell{c}{Number of equally powered miners}\\
      \hline
    BTC & 7.81 - 9.16\\
     \hline
    BCH & 8.03 - 8.71\\
     \hline
    BSV & 4.39 - 4.47\\
     \hline
    ETH & 7.80 - 11.58\\
     \hline
    ZEC & 11.78 - 13.79\\
     \hline
    LTC & 3.52 - 3.59\\
 \hline
\end{tabular}
  \caption{The number of equally powered miners that would have the same entropy as that cryptocurrency's mining power
    distribution.}
  \label{tab:entropy}
\end{table}

Blocks that were not claimed by any miner introduce a layer of uncertainty in this computation.
To handle these cases, we first assume that unclaimed blocks do not belong to miners that have claimed other blocks as
    their own.
Then, the pessimistic case is that the unclaimed blocks were all mined by a single miner, which gives us a lower bound
    on $n$.
Similarly, to get an upper bound on $n$, we treat all unclaimed blocks to be mined by distinct miners.
We present the lower and upper bound of $n$ for all coins in our study in Table~\ref{tab:entropy}.

\textbf{Results.}
Similar to previous studies~\cite{adem18}, we observe that none of these coins are very decentralized.
A consensus protocol with fourteen equally powerful, independent members would be more decentralized than the most
    decentralized cryptocurrency, ZEC.
However, even among this small range, there are fairly substantial differences in decentralization across
cryptocurrencies.
The most centralized coins, LTC and BSV, have the equivalent of four to five equally powerful miners controlling their cryptocurrency.
This is two to three times more centralized than the most decentralized coin, ZEC, which has about twelve to fourteen
    such miners.

\subsection{Existence of Previous Block Advantage}
\label{sec:selfmining}

We now ask the question of whether the hypothesized previous block advantage exists in our dataset.
To do this, we compute the hashpower of a miner $M$ in two different ways.
First, we use the proportion of blocks that were mined by $M$ in the entire sample period.
Then, we use the proportion of blocks that were mined by $M$, but only when we look at blocks that were mined
    immediately after $M$ mines a block.
If the previous block advantage does not exist, then both sample proportions should be equal, which is our null
    hypothesis.
We run a two proportion z-test to see if this is indeed the case for each of the miners in the six cryptocurrencies
    we studied.
Note that the two samples are independent because the miner of any particular block on the chain is chosen independently
    at random from the set of all miners.

\begin{table*}[htbp]
\center
\setlength{\tabcolsep}{0.1mm}{
  \begin{tabular}{|c|ccccccccc|}
    \hline
  	\multirow{2} * {BTC}&BTC.com&AntPool&SlushPool&ViaBTC&BTC.TOP&F2Pool&BitClub Network&BitFury&BTCC Pool\\
  	\cline{2-10}
& \cellcolor[gray]{.8} 0.0083&0.34&0.86&0.66&0.8&0.28&0.14&0.99&0.98\\
    \hline
\multirow{2} * {BCH}
&CoinGeek&BTC.TOP&BTC.com&BMG Pool&ViaBTC&Bitcoin.com&AntPool&Rawpool&\\
\cline{2-10}
& \cellcolor[gray]{.8} 2.7e-07& \cellcolor[gray]{.8} 1.2e-89&0.96& \cellcolor[gray]{.8} 0.0065&  0.16& \cellcolor[gray]{.8} 0.021& 0.38& \cellcolor[gray]{.8} 0.0014&\\
    \hline
\multirow{2} *{BSV}
&CoinGeek&svpool.com&Mempool.com&ViaBTC&BMG Pool&Poolin&&&\\
\cline{2-10}
& \cellcolor[gray]{.8} 2.4e-130& \cellcolor[gray]{.8} 1.3e-33& \cellcolor[gray]{.8} 4.3e-20& \cellcolor[gray]{.8} 1.6e-11& \cellcolor[gray]{.8} 1.1e-17& \cellcolor[gray]{.8} 6.4e-12&&&\\
    \hline
\multirow{4} *{ETH}&Ethermine&SparkPool&F2Pool\_2&Nanopool&MiningPoolHub\_1&BitClub Pool&DwarfPool\_1&bw&miner0\\
\cline{2-10}
& \cellcolor[gray]{.8} 2.1e-18& \cellcolor[gray]{.8} 3.3e-14& \cellcolor[gray]{.8} 1.4e-12& \cellcolor[gray]{.8} 1.7e-07& \cellcolor[gray]{.8} 1.4e-07& \cellcolor[gray]{.8} 2.7e-07& \cellcolor[gray]{.8} 0.034&  0.1& \cellcolor[gray]{.8} 1.8e-05\\
\cline{2-10}
&miner10&miner2&&&&&&&\\
&0.54& \cellcolor[gray]{.8} 0.038&&&&&&&\\
    \hline
\multirow{4} *{ZEC} &miner12&miner3&miner2&miner61&miner1&miner14&miner19&miner6&miner7\\
\cline{2-10}
&  0.2& 0.35&0.79& \cellcolor[gray]{.8} 3.3e-14&0.82&0.64& 0.41& 0.092&  0.15\\
\cline{2-10}
&miner4&miner5&miner13&miner0&miner8&miner24&&&\\
\cline{2-10}
& 0.13& \cellcolor[gray]{.8} 2.7e-20& \cellcolor[gray]{.8} 0.039& \cellcolor[gray]{.8} 0.00037&1& 0.23&&&\\
\hline
\multirow{2} *{LTC}&AntPool&ViaBTC&LTC.TOP&\tabincell{c}{litecoinpool.org}&&&&&\\
\cline{2-10}
& 0.077&0.83& \cellcolor[gray]{.8} 1.4e-24& \cellcolor[gray]{.8} 8.3e-14&&&&&\\
    \hline
  \end{tabular}
  }
    \caption{These are p-values from the two-proportion z test whose null hypothesis is that the two sample mining power estimates
    are the same, which would disprove the existence of the previous block advantage. We use a $5\%$ significance
    threshold and highlight, in gray, all values that are statistically significant.}
  \label{tab:p-value}
\end{table*}

Table~\ref{tab:p-value} shows the p-values from the two proportion z test with statistically significant results 
    shaded.
As our data is fairly noisy, we are more likely to make a type two error so we use a significance level of $5\%$.
Note that a Bonferroni correction~\cite{bonferroni} is not required as each of our proportions are drawn from an
    independent set of blocks so the result of one of our z tests has no effect on the result of other tests.

\textbf{Results.}
The previous block advantage is present across many small and large miners, with $49\%$ of miners across all
    cryptocurrencies obtaining a statistically significant advantage.
If we consider just the largest four miners in each cryptocurrency, $62.5\%$ have the previous block advantage, which
    shows the prevalence of this advantage amongst larger miners.

Some cryptocurrencies have a stronger previous block advantage than others.
For example, Bitcoin (BTC) and ZCash (ZEC) have very few advantaged miners, which may have to do with the design
    of the corresponding network layers.
Bitcoin has multiple relay networks~\cite{adem18,fibre} that help blocks propagate faster and alleviate the
    previous block advantage.
ZCash's mining distribution is the most decentralized out of any cryptocurrency, which is a potential reason for ZEC's
    low previous block advantage.
In contrast, we see that Bitcoin Cash (BCH), Bitcoin SV (BSV), and Ethereum (ETH) has most, if not all, of the top
    miners benefitting from a statistically significant previous block advantage.

The previous block advantage in ETH seems to extend to almost every miner including smaller miners.
This supports the existence of centralization pressure in ETH due to the previous block advantage.
The likely explanation behind the prevalence of the previous block advantage in ETH is due to the comparatively
    smaller block interval in ETH.
This makes it easier for our statistical tests to detect the existence of the previous block advantage.
The failure to detect this in other cryptocurrencies is likely due to the larger block interval causing a smaller
    sample size.
All other metrics measuring the prevalence of the previous block advantage do not show that Ethereum is particularly
    facing this centralization pressure.
As a result, we believe that the previous block advantage is inherent to proof of work mining.

Note that some standard mining practices, such as spy mining, which occurs when a mining pool adds some of their 
    mining power to another pool to obtain block headers more quickly, will not increase the previous block advantage.
These practices are a way for miners to reduce the block transmission time to the network, which reduces the previous
    block advantage.

\textbf{Caveats.} 
Our data, especially among smaller miners, is fairly noisy since they have mined comparatively few blocks.
Thus, making statistical inferences is more difficult as the previous block advantage obtained by them would have to be
    quite large.
However, our results strongly suggest that the previous block advantage is real and present in our dataset.

\textbf{Succession Matrix.}
Since we know that the previous block advantage exists, we systematically study these network effects using the
    \emph{succession matrix} of a cryptocurrency.
To build the succession matrix, index each known miner in the cryptocurrency as $M_1, M_2, \dots M_n$ where 
    $M_i$ has more hashpower than $M_j$ if $i > j$ based on the estimate from the full blockchain.
$S_{ij}$ is computed as the proportion of blocks that $M_j$ has mined when considering only the blocks immediately
    following blocks mined by miner $M_i$.
This matrix allows us to look at the previous block advantage that miners obtain when mining after each other as well
    as themselves.
We include the succession matrices for all studied cryptocurrencies in Figure~\ref{fig:freq_p} 
    located in Appendix~\ref{app:fig}.

\subsection{Measuring the Previous Block Advantage}
\label{sec:distmetrics}

A metric to quantify the prevalence of the previous block advantage is important as it allows us to compare how
    prevalent the advantage is across cryptocurrencies as well as across the same cryptocurrency over time.
To reason about how to construct such a metric, we start with the succession matrix.
If miner $M_j$ gets a large previous block advantage when some miner $M_i$ mines the previous block, then we would
    expect the entry $S_{ij}$ to be large relative to the mining power of $M_j$.
Thus, our first step to construct this metric is to normalize the succession matrix to highlight how large the previous
    block advantage is.

The normalized succession matrix $N$ is the matrix where $N_{ij} = \frac{S_{ij}}{m_j}$, where $m_j$ is the mining power
    of miner $M_j$.
If $N_{ij}$ is larger than one, then $M_j$ has a slight advantage when mining after $M_i$ mines a block and if $N_{ij}$
    is smaller than one, $M_j$ is at a disadvantage.
We include the normalized succession matrix for all studied cryptocurrencies in Figure~\ref{fig:freq_d} 
    located in Appendix~\ref{app:fig}.

A first attempt at such a metric might simply be to compare $N_{ii}$ to $1$ and add them up across all miners, i.e. 
    $\sum_{i=1}^n (N_{ii} - 1)$ where $n$ is the number of miners.
However, that metric treats small and large miners equally, which inaccurately describes the prevalence of the previous
    block advantage.
A miner gets the previous block advantage every time they mine a block, so the more hashpower a miner has, the more
    heavily they should be weighted.
With that intuition, if we let $P_i$ be the fraction of hashpower that $M_i$ controls, our final distance
    metric is $ D = \sum_{i=1}^{n} (N_{ii} - 1) P_i$.
A system with no net previous block advantage will have $D \leq 0$.

\textbf{Results.}
Table~\ref{tab:diagonal_value} shows the resulting distance metrics for all cryptocurrencies in our study.
BTC miners have a very small previous block advantage compared to other cryptocurrencies, which is likely
    due to the small block size and the presence of multiple fast relay networks that improve block
    propagation.
Interestingly, ETH has a fairly low previous block advantage metric even though most ETH miners have a statistically
    significant previous block advantage.
This lends credence to the theory that the previous block advantage was more easily detectable
    in ETH due to the smaller block interval.
Finally, we see that outliers have a large effect on these metrics.
While many BCH miners do not obtain a significant previous block advantage, our metric is large for BCH
    due to the large previous block advantage obtained by BTC.TOP, which is one of BCH's largest miners.

\begin{table}
\center
  \begin{tabular}{|c|c|c|}
      \hline
      Coin & \tabincell{c}{Previous Block\\Advantage (Biased)} & \tabincell{c}{Previous Block\\Advantage} \\
      \hline
 BTC & $-0.003 \pm 0.003$ & -0.001\\
BCH & $0.372 \pm 0.0308$ & 0.49\\
BSV & $1.04 \pm 0.135$ & 0.249\\
ETH & $0.104 \pm 0.003$ & 0.103\\
ZEC & $0.242 \pm 0.0308$ & 0.264\\
LTC & $0.265 \pm 0.025$ & 0.318\\
      \hline
  \end{tabular}
    \caption{These are the metrics that describe how prevalent the previous block advantage is in each cryptocurrency.
    The first column uses the mining power distribution implied by looking at the blockchain and the second column uses
    a debiased mining power distribution which we discuss in more detail in Section~\ref{sec:debias}.}
  \label{tab:diagonal_value}
\end{table}

On further investigation, the reason for this outlier is that BTC.TOP does active chain switching between
    BCH and BTC.
Thus, BTC.TOP gets a previous block disadvantage of sorts: when BTC.TOP has not mined the previous block, it is
    less likely to be mining on BCH and is consequently less likely to mine the next block.
This behavior violates our assumption that the hashpower of each miner remains constant throughout the measurement
    period.
BTC.TOP is not as influential of a miner in BTC compared to BCH and consequently does not skew the conclusions we make for BTC as much.
We provide further evidence for BTC.TOP's behavior in Section~\ref{sec:btctop}.

\begin{figure*}[htbp]
\centering
\includegraphics[width=\linewidth]{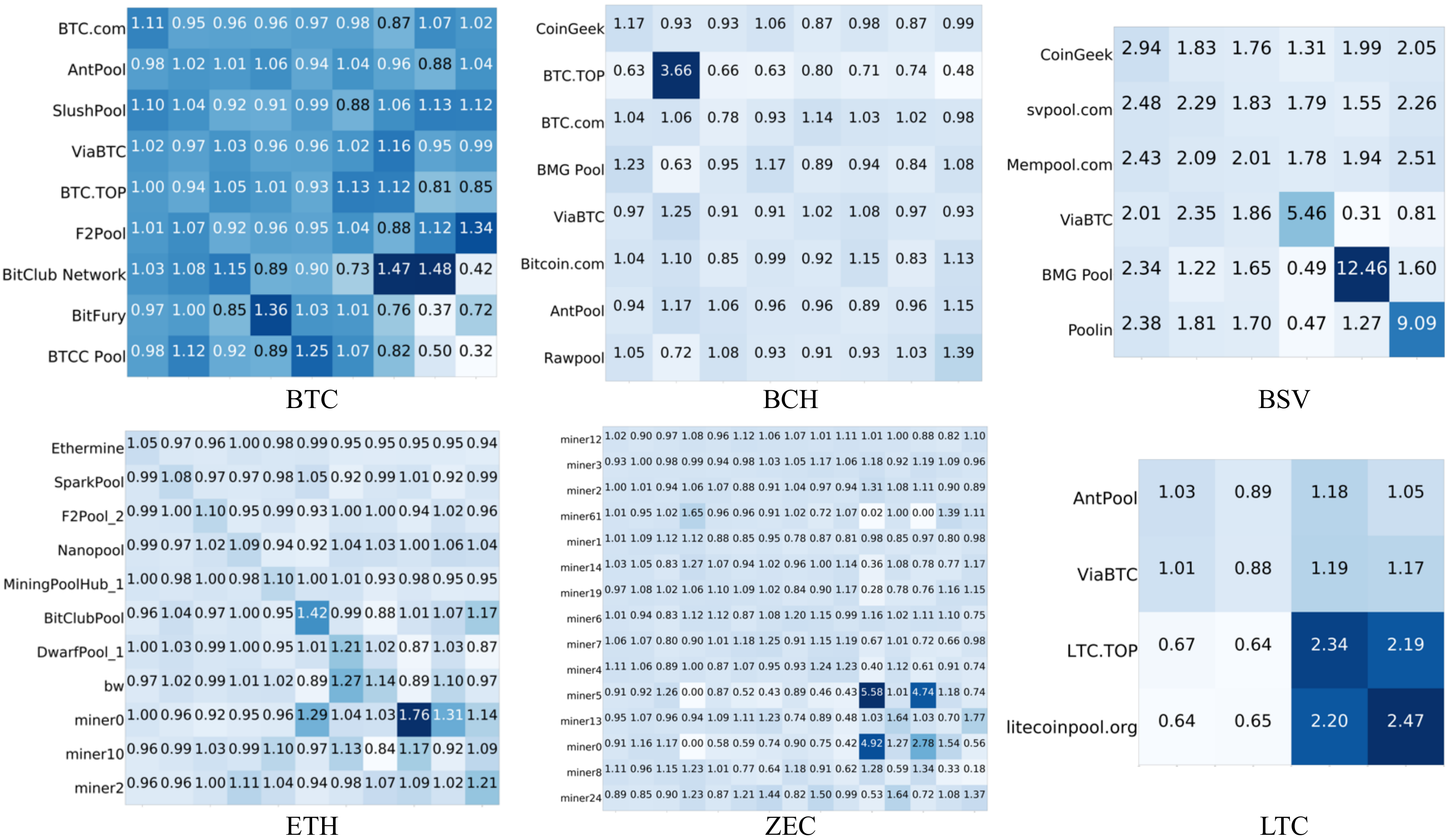}
\caption{These are the normalized succession matrices for all cryptocurrencies when using the 
    debiased mining power distribution to normalize the succession matrix.}
\label{fig:freq_real_eth}
\end{figure*}

\subsection{Correcting for Previous Block Advantage}
\label{sec:debias}

The proportion of blocks belonging to a miner on the main chain is typically used to estimate their
    hashpower~\cite{adem18,blockchaininfo}.
However, the previous block advantage skews these estimates as larger miners have a larger advantage.
Understanding the hashpower distribution of each cryptocurrency is critical as miners who get too powerful
    can be put under pressure to scale back their operations, such as when a miner obtained 51\% of the hashpower
    in Bitcoin in July 2014~\cite{ghashissue2014}.
In this section, we show how to correct for this effect in order to more accurately estimate the mining power
    distribution.

Denote the $n$ miners of a cryptocurrency as $M_1, M_2, \cdots, M_n$ and the true, unbiased proportion of the
    mining power owned by miner $M_i$ is $O_i$.
Note that $\sum_{i = 1}^n O_i = 1$, and that $O_i$ is the probability that $M_i$ has mined a randomly
    selected block in the blockchain.

Let $\alpha_i$ be the relative amount that $M_i$'s hashpower is increased due to its previous block advantage.
Thus, $M_i$ has a relative probability of $O_i + \alpha_i$ of mining the next block when it as mined the
    previous block.
Any other miner, $M_j$, just has a relative probability of $O_j$ of mining the next block.
To get the absolute probability that any miner has mined a block, we renormalize the probabilities by dividing
    by $1+ \alpha_i$.
Thus, we see that the probability that $M_i$ mines a block immediately after mining the previous block is $\frac{O_i +
\alpha_i}{1 + \alpha_i}$.
This is also the diagonal entry $S_{ii}$ in the successor matrix.

We now find an expression for the fraction of blocks mined by $M_i$ on the main chain using $O_i$ and
    $\alpha_i$.
Since $M_i$ has $O_i$ of the hashpower, $M_i$ has a probability of $O_i$ to mine a particular block.
However, $O_i$ of the time, $M_i$ gets an additional $\alpha_i$ probability to mine the next block.
Thus, we see that the relative probability of $M_i$ mining any block on the main chain is $O_i(1+ \alpha_i)$.
To renormalize, we divide the relative probability by $\sum_{j=1}^{n} O_i(1+\alpha_j)$.
Thus, the proportion of blocks mined by $M_i$ on the main chain is $\frac{O_i(1+\alpha_i)}{1 + \sum_{j=1}^{n}
O_j\alpha_j}$.

This is the proportion of blocks that were mined by $M_i$ on the blockchain or the biased mining power
    estimate.
We can solve the resulting set of $2n$ linear equations in order to solve for $O_i$, as desired.

\begin{table*}[hbtp]
\setlength{\tabcolsep}{0.3mm}{
  \begin{tabular}{|c|ccccccccl|}
      \hline
  \multirow{3} * {BTC} &BTC.com&AntPool&SlushPool&ViaBTC&BTC.TOP&F2Pool&BitClub Network&BitFury&BTCC Pool\\
  \cline{2-10}
  ~ & 0.2568 & 0.1699 & 0.1416 & 0.1347 & 0.1248 & 0.1090 & 0.0260 & 0.0207 & 0.0161\\
~&0.2498&0.1701&0.1441&0.1362&0.1270&0.1090&0.0259&0.0211&0.0164
\\
      \hline
  \multirow{3} * {BCH}  & CoinGeek & BTC.TOP& BTC.com&BMG Pool&ViaBTC& Bitcoin.com& AntPool& Rawpool&\\
  \cline{2-10}
  ~ & 0.2453 & 0.1554 & 0.1163 & 0.1090 & 0.1090 & 0.0979 & 0.0931 & 0.0736 &\\
  ~ &0.2486&0.1112&0.1288 &0.1144 &0.1166 &0.1033 &0.1003 &0.0764  &\\
      \hline
  \multirow{3} * {BSV}  &CoinGeek&svpool.com&Mempool.com&ViaBTC&BMG Pool&Poolin&&&\\
  \cline{2-10}
  ~ &0.2169&0.1262&0.109&0.03278&0.02355&0.02051&&&\\
  ~ &0.1581&0.1195&0.1096&0.03205&0.02053&0.01971&&&\\
      \hline
    \multirow{6} * {ETH}  & Ethermine&SparkPool&F2Pool\_2&Nanopool&MiningPoolHub\_1&BitClub Pool&DwarfPool\_1&bw&miner0\\
      \cline{2-10}
    ~& 0.3202 & 0.1867 & 0.1503 & 0.1191 & 0.1015 & 0.0332 & 0.0220 & 0.0181 & 0.0168 \\
~&0.3172&0.1866&0.1505&0.1200&0.1022&0.0334&0.0223&0.0184&0.0169\\
 \cline{2-10}
~&miner10&miner2&&&&&&&\\
\cline{2-10}
~& 0.0166 & 0.0148&&&&&&&\\
~&0.0169&0.0150&&&&&&&\\
      \hline
  \multirow{6} * {ZEC}  & miner12	&miner3&miner2&miner61&miner1&miner14&miner19&miner6&miner7	\\
    \cline{2-10}
  ~ & 0.1922 & 0.1906 & 0.1353 & 0.0992 & 0.0522 & 0.0515 & 0.0495 & 0.0455 & 0.0441  \\
  ~ &0.1936 &0.1929 &0.1382 &0.0936 &0.0532 &0.0523 &0.0500 &0.0456 &0.0443\\
  \cline{2-10}
  	&miner4	&miner5&miner13&miner0&miner8	&miner24&&&\\
  	  \cline{2-10}
  & 0.0346 & 0.0309& 0.0229 & 0.0205 & 0.0157 & 0.0146 &&&\\
   &0.0347 &0.0272 &0.0228 &0.0200 &0.0160 &0.0147 &&&\\
      \hline
  \multirow{3} * {LTC} & AntPool & ViaBTC & LTC.TOP & \tabincell{c}{litecoinpool.org}&&&&&\\
  \cline{2-10}
  ~& 0.4282 & 0.3241 & 0.1424 & 0.1051 &&&&&\\
  ~ &0.4324 &0.3556 &0.1199 &0.0920  &&&&&\\
      \hline
\end{tabular}
}
  \caption{We present the mining power distribution for each cryptocurrency after removing the effect of the previous
    block advantage. The first row is the identities of each miner. The second row is the biased estimate using the raw
    blockchain sample. The third row is the debiased estimate.
}
  \label{tab:real-mining-power}
\end{table*}

\textbf{Results.}
Table~\ref{tab:real-mining-power} shows the biased and corrected estimates for the mining power distribution.
As we expect, since large miners have a bigger previous block advantage, their mining power was overestimated when
    using solely the proportion of blocks on the chain.
Using the modified mining power distribution, we recompute the normalized succession matrix.
Figure~\ref{fig:freq_real_eth} shows all normalized succession matrices and the distance metrics introduced in Section~\ref{sec:distmetrics} 
    are shown in Table~\ref{tab:diagonal_value}.
The recomputed values indicate that there is a larger previous block advantage than we initially believed.

Small discrepancies in the mining power distribution matter immensely to individual miners and underutilizing their
    hashpower by even a tiny amount might equate to tens of thousands of dollars lost.
An accurate estimate of a miner's hashpower and how advantaged they are in the network allows miners to understand how 
    to improve their mining operations.

\textbf{Validation.}
As forked blocks are discarded and forgotten in most cryptocurencies, this corrected mining power distribution is not
    possible to validate.
However, Ethereum's forked blocks, called uncle blocks, are stored and rewarded in accordance with the GHOST~\cite{ghost} protocol.
Uncle blocks can be stored on the chain much later than when they are mined, which eliminates the previous block
    advantage as the propagation of uncle blocks can be much slower than regular blocks.
Therefore, including uncle blocks when computing the mining power distribution will also eliminate the previous block
    advantage.
We include a detailed comparison of these two distributions in Table~\ref{tab:real-mining-power-uncle} in
    Appendix~\ref{app:fig}.
The high level takeaway is that our technique provides a more accurate estimate of the mining power distribution than
    simply using blocks on the main chain, with discrepancies between the estimates being attributable to noise.

\section{Block Propagation}
Understanding block propagation is key to resolving many core debates in cryptocurrencies, from setting parameters like
    the block size to deciding which parts of the codebase need to be optimized for maximum impact.
Using our techniques, we can provide insights on how blocks propagate through the network and help cryptocurrencies make
    data-driven decisions.

In this section, we first estimate the latency of block propagation in each network using the succession matrix.
Then, we introduce and discuss the safe envelope of a cryptocurrency in order to show how the previous block advantage
    changes with respect to block size and how to evaluate changes to the network.
    The decision, such as a network upgrade, actually achieved the desired effect.

\subsection{Latency of Block Propagation}
Block propagation speed affects many aspects of decentralization, such as
    the prevalence of the previous block advantage and how often forks occur.
We can use the succession matrix to estimate how long blocks take to travel from one miner to all other miners in each
    cryptocurrency.

We can use a simplified model of block propagation to compute the average latency, where miner $M_i$ mines 
    a block that takes $L$ minutes to propagate to all other miners.
If the average block interval is $t$, then the miner $M_i$ effectively increases its proportion of the mining power
    by a factor of $\frac{L}{t}$ every time it mines.
This increase can also be computed from the succession matrix, $S$, as well.
If $M_i$ mines after itself $S_{ii}$ of the time and has $O_i$ fraction of the hashpower, its 
    advantage is $\frac{S_{ii} - O_i}{O_i}$.
To find the average latency, $\bar{L}$, we take the average advantage over all $n$ miners:

\[\frac{\bar{L}}{t} = \sum_i^{n} \frac{S_{ii} - O_i}{O_i} O_i = \sum_i^{n} S_{ii} - 1\]

Thus, the average latency is $\bar{L} = ( (\sum_i^{n} S_{ii}) - 1)t$.

\begin{table}
\center
  \begin{tabular}{|c|c|c|c|}
      \hline
    Coin & Avg Block Interval (min) & Latency (min) & 95\% conf \\
      \hline
      BTC & 9.50 & -0.01 & (-1.38, 1.35)\\
      BCH & 7.20 &  2.68 & (1.54, 3.83)\\
      BSV & 9.97 & 5.68 & (3.59, 7.77)\\
      ETH & 0.24 & 0.02 & (0.01, 0.04)\\
      ZEC & 2.40 & 0.58 & (0.07, 1.09)\\
      LTC & 2.50 & 0.66 & (0.41, 0.92)\\
      \hline
\end{tabular}
    \caption{We present the latency to transmit a block between miners for different coins. The columns represent,
    respectively, the average time between blocks, the latency to transmit blocks, and the 95\% confidence interval of
    that latency.}
  \label{tab:latency}
\end{table}

\begin{figure*}
\centering
\includegraphics[width=\linewidth]{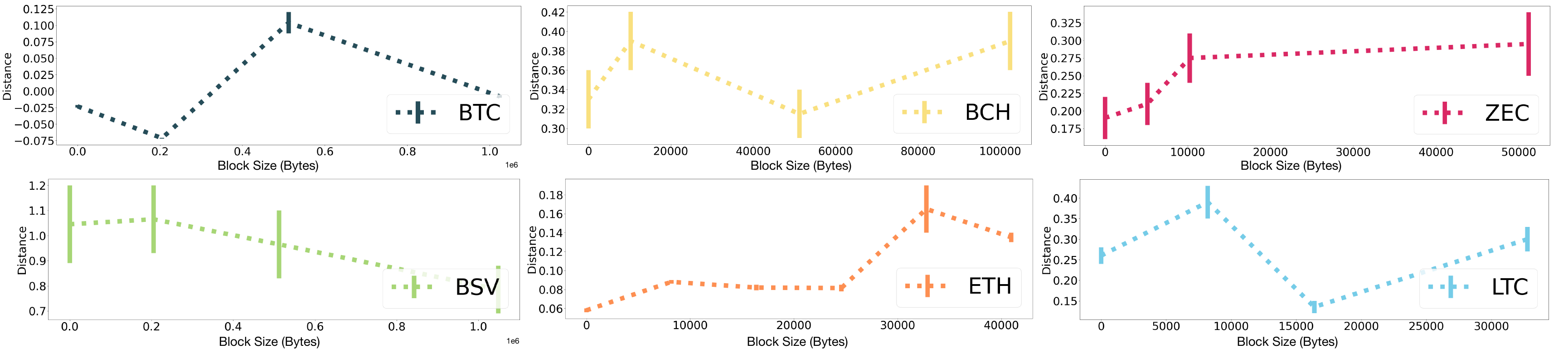}
\caption{This graph shows the prevalence of the previous block advantage of three coins compared to block size.
    A point in this graph represents the interval of sizes from that point up to the next one.
    For example, the point at 0 in BTC's line consists of all blocks from 0 to 204800 bytes.
    The dotted line is used to estimate which interval the safe envelope for each cryptocurrency lies.}
\label{fig:self-mining-graph}
\end{figure*}

\textbf{Results.}
Table~\ref{tab:latency} presents the latency of each cryptocurrency in minutes and the $95\%$ confidence interval.
The average latency of BTC is particularly small, with the midpoint of the confidence interval being slightly negative.
Most other coins have a block latency that is about $10$ to $30$ percent of their average block interval time.
Thus, it would be useful for many coins to focus on improving their block propagation characteristics.
BCH's latency estimate is also unnaturally high, and we discuss the reasons why below.

\textbf{Sanity Check.}
To check our latency results, we use the orphan rate, which can be approximated as $1 - e^{\frac{-\tau}{T}}$, where
    $\tau$ is the average latency between miners and $T$ is the block interval time~\cite{Rizun2016}.
From prior measurement work in the Bitcoin (BTC) and Ethereum (ETH) networks, we see that the orphan rate
    for Bitcoin is extremely small and Ethereum's fluctuates between 6-10\%~\cite{adem18}.
From our table above, we see that the orphan rate in BTC is very close to zero which is in line with prior work.
In ETH, the 95\% confidence interval for the orphan rate is between 4\% and 15\%, also in line with prior measurements.

\textbf{Caveats.}
The largest caveat for our measurement comes from the accuracy of the timestamp data that was
    collected from each block.
Since those are set by the miner, they are subject to manipulation either due to benign reasons, 
    such as clock skew, or malicious actions by miners.
To validate the timestamp data, we compared them with the receive time in the block explorers and 
    checked the timestamps of some blocks on our full nodes for Bitcoin and Ethereum.

Our latency measurement is sensitive to outliers, which is why BCH's latency number is quite high at 2.68
    minutes.
The outlier in the BCH data is the inflated previous block advantage that BTC.TOP has due to its unusual mining
    behavior.
    
\begin{figure}
\centering
\includegraphics[width=\linewidth]{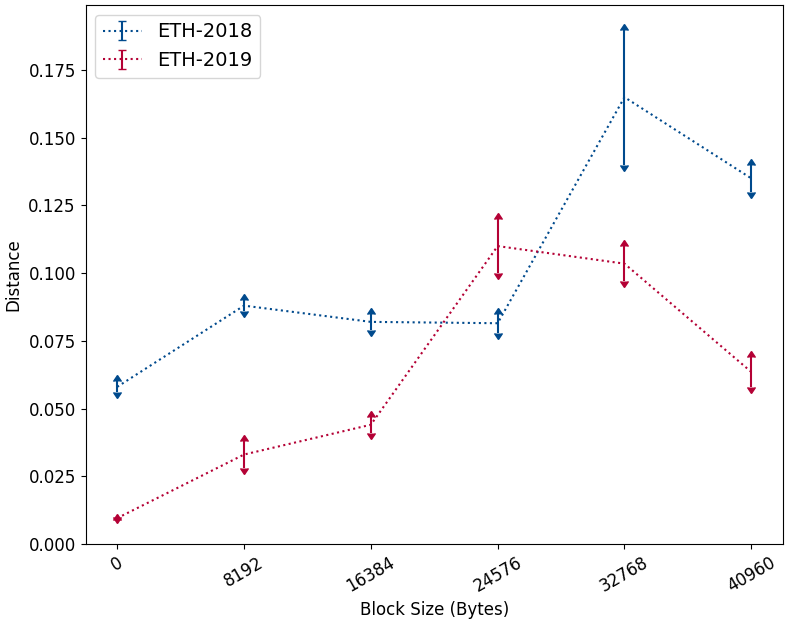}
\caption{This graph compares the prevalence of the previous block advantage across two snapshots of Ethereum
    that are separated by a client upgrade improving block propagation.}
\label{fig:self-mining-graph-eth}
\end{figure}

\subsection{Safe Envelopes}
Understanding how the block size affects decentralization has been a critical debate in cryptocurrencies, is
   the root cause behind some chain splits.
To get at this question, we use the previous block advantage metric from Section~\ref{sec:distmetrics} and compute
   it on blocks after grouping them by size.


We expect to see that the previous block advantage is small for small blocks since small
    blocks take less time to propagate.
However, as the block size increases, the previous block advantage would get larger.
We define the \emph{safe envelope} of a cryptocurrency as the last block size range that does 
    not have a significantly higher previous block advantage relative to a block with no payload.
Beyond the safe envelope, increasing the block size directly encourages miners to centralize
    as there is an increased previous block advantage.

\textbf{Results.}
Figure~\ref{fig:self-mining-graph} presents the previous block advantage for different block sizes.
Error bars represent the 95\% confidence interval for each block size.
From the figure, we see that Bitcoin (BTC), Bitcoin Cash (BCH) and Litecoin (LTC) have a safe envelope that is beyond
    the block sizes that we have observed.
In Bitcoin (BTC), the range is likely due to a maximum block size that is too small for the network.
However, Litecoin and Bitcoin Cash blocks are not yet at their maximum block size, so the lack of a defined safe
    envelope stems from underutilization not the block size limit.
This means that BTC can safely increase its block size, while LTC and BCH users are free to utilize the chain
    more without risking mining centralization.

On the other hand, we see that ZCash, Ethereum, and Bitcoin SV all have a defined safe envelope.
This means that they all operate at a throughput that encourages some mining centralization.
ZCash has a safe envelope that ends somewhere between 10 KB and 50 KB while Ethereum's safe 
    envelope ends somewhere between 32 KB and 40 KB.
Bitcoin SV's safe envelope ends somewhere after 10 KB.
These are extremely coarse-grained estimates, which are limited by the number of blocks in our sample.

Using the safe envelope, we can analyze the effects of significant events through a concrete example.
In December 2018, Ethereum released a new client that improved the block propagation time on the
    Ethereum network.

Figure~\ref{fig:self-mining-graph-eth} compares two different Ethereum snapshots: one from just before the
    deployment and one from after the deployment when most users had upgraded.
Interestingly, we see that improving the network layer does not always improve the safe envelope.
Indeed, the safe envelope in 2018 was between 24 KB and 32 KB while the safe envelope in 2019 actually
    decreased to somewhere between 16 KB and 24 KB.
However, if we look at the magnitude of our distance metric, it has decreased for almost
    all block sizes.
We therefore conclude that the network upgrade improved decentralization in the system as a whole even
    though the safe envelope has decreased.

\section{Unusual Miner Behaviors}
Using our analysis techniques, we have observed some miner cartels in ZEC, BSV and LTC. Additionally, we have observed
   BTC.TOP switching between BCH and BTC. We discuss how we reached those conclusions and the methodology behind them
   in Appendix~\ref{app:unusual}.

\section{Related Work}
Cryptocurrency measurement has been an active research area for many years, starting with the work by Decker et
    al~\cite{DeckerW13}, which connected to many peers in the Bitcoin network and measured block propagation
    characteristics.
Since then, traditional cryptocurrency measurement studies have typically chosen a few questions to ask, built a
    corresponding probe and then inferred properties of the peer to peer network based on measurements made by their
    probe.
AddressProbe~\cite{Miller2015DiscoveringB} discovered peer-to-peer links in Bitcoin to analyze the topology of the
entire network, find connected components, and find high degree nodes.
The Falcon Relay Network and BMS~\cite{adem18} was a blockchain measurement system to measure the network properties 
   of the nodes in the Bitcoin and Ethereum peer to peer networks.
TxProbe~\cite{SSCJAAB18} cleverly used orphaned transactions in order to take snapshots of Bitcoin's testnet.
NodeFinder~\cite{Kim2018MeasuringEN} measured the Ethereum network propagation characteristics in the peer to peer network.
Still, other measurement studies primarily focused on Bitcoin~\cite{AGVS2014,Pappalardo2018BlockchainII,AJ2015} or
Ethereum~\cite{Kiffer2018AnalyzingEC} through direct measurement.

However, while these techniques are very powerful in their own right, they do have a few limitations.
Prior techniques rely on a snapshot of the network during a particular time period- it is impossible to go back in time
    and answer historical queries.
It is also very difficult to effectively capture the effects of significant events, such as an Ethereum network upgrade,
    since the measurement probe must be running during the time of the event.
Additionally, measurement probes require significant manpower to build, run, and maintain.
Finally, measurement probes are fundamentally limited in what types of questions they can answer as they can only be run
    on nodes that are controlled by the researcher.
Thus, gaining insights, such as the structure of the mining network, is very difficult to do with a traditional
    measurement probe.

Our large scale statistical analysis relies on infrastructure that has already been built and maintained:
    the codebase of the cryptocurrency itself.
Leveraging this data is very powerful and sidesteps all of the problems present in building custom measurement probes.
The codebase is always running and is running on every single node in the network, which allows us to collect
    information from the entire network at all times.
However, our technique does require a significant amount of data to make meaningful conclusions and the estimates are
    more coarse-grained than they would be with a measurement probe.

Prior work also verifies our conclusions as well.
For example, we are not the first paper to note the centralization of the mining distribution~\cite{adem18}.
Other works~\cite{AGVS2014,AJ2015} show recent events
in Bitcoin are showing the limits of its decentralized nature and how Bitcoin is slowly becoming centralized.
Our analysis on block propagation also suggests that the Bitcoin is underutilizing its bandwidth and that the network
   can use more bandwidth without compromising decentralization, a conclusion that has been made using prior measurement
   probes as well~\cite{adem18,onscalingdecentralized}.

While many measurement studies have focused on the largest cryptocurrencies, Bitcoin and Ethereum, there has been some
    prior work looking at properties of other cryptocurrencies as well.
For example, some work has done a comparative financial and statistical analysis between Bitcoin, Litecoin, Ripple and Ethereum~\cite{Sapuric2017LedgerDW}.
Additionally, others have compared Bitcoin, Ethereum and Ripple on their architecture, scripting language and other crucial
    properties~\cite{Mauri2018ACA}.
There have also been in-depth studies on comparing the different consensus mechanisms~\cite{Bonneau2015SoKRP,Bonneau2015perspectivesOB}.
However, to our knowledge, we are the first known measurement study incorporating six different cryptocurrencies.

In our work, we looked at the safe envelope in order to deduce whether or not it was safe to increase the block size
    in each cryptocurrency.
However, other, more effective ways of scaling cryptocurrencies have been proposed as well which optimize all parts of
    the cryptocurrency stack.
Starting at the bottom, the peer to peer network has many suggested optimizations that make the block and transaction 
    propagation more efficient~\cite{NaumenkosErlay2019paper,graphene}.
In the consensus layer, many consensus protocol changes have been proposed for Bitcoin that keep the same security
    guarantees as Nakamoto consensus while increasing the throughput of the system~\cite{eyal2016bitcoin,byzcoin}.
Finally, other proposals look at removing transactions from the blockchain altogether and only use the blockchain as
    a conflict resolution mechanism~\cite{lightning,plasma}.
In all of these proposals, however, measuring the safe envelope is critically important so that there is empirical
    evidence whether or not the new technology actually made a meaningful change to the underlying system.



\section{Conclusion}
This paper rests on a single, crucial observation: mining a block gives the corresponding miner a
    slight, but measurable advantage when mining the next block.
Leveraging this insight is extremely powerful and allows us to answer difficult to answer questions
    about the internals of the network that connects miners.
This, in turn, allows us to provide actionable intelligence to miners and developers to see where
    they should spend their efforts in order to achieve their objectives.
With these techniques, we were able to compute a more accurate mining power distribution, answer
    questions about block propagation including how block size affects the previous block advantage,
    and observe some interesting miner behaviors like miner cartelization and active chain switching.

%
%
%

%

\bibliographystyle{splncs04}
\bibliography{main}

\appendix

\section{Unusual Miner Behaviors}
\label{app:unusual}
We now present two interesting miner behaviors present in our dataset.
First, we discuss miner cartelization when two miners are interconnected and essentially
    function as a single miner.
Next, we discuss the behavior of BTC.TOP and active chain switching.

\subsection{Miner Cartelization}
Miner cartels are groups of miners that are nominally separate, but act like a single entity 
    separate from the rest of the mining network.
Such miners would preferentially send their blocks to each other and may have dedicated links 
    between each other.
When such behavior happens, miners in the cartel receive blocks from each other before everyone 
    else.
Miner cartels are dangerous because they make the cryptocurrency appear more decentralized than it
    actually is.
Detecting cartels allows the community to understand, and perhaps take action, if the
   cryptocurrency is becoming too centralized.

Miners in a cartel receive a previous block advantage anytime another miner in the same cartel mines
   a block.
Thus, we would expect to see very high values for miners in a cartel in the normalized succession
    matrix.
To highlight cartel miners more clearly, we modify the normalized succession matrix, $N$, as follows.

Suppose we have two miners, $M_i$ and $M_j$, that belong to the same cartel.
That means that $M_i$ gets an advantage whenever $M_j$ mines a block, and vice versa.
Consequently, $N_{ij}$ and $N_{ji}$ must both be much larger than one, so we can analyze pairs of miners
    in the same cartel by computing $\frac{N_{ij} + N_{ji}}{2} - 1$.
High positive values imply that $M_i$ and $M_j$ may be in the same cartel.
We present this in a pairs matrix, $P$, where $P_{ij} = P_{ji} = \frac{N_{ij} + N_{ji}}{2} - 1$.
If there were no miner cartels, then we would expect $P$ to contain solely elements very close to $0$.

\begin{figure*}
\centering
\includegraphics[width=\linewidth]{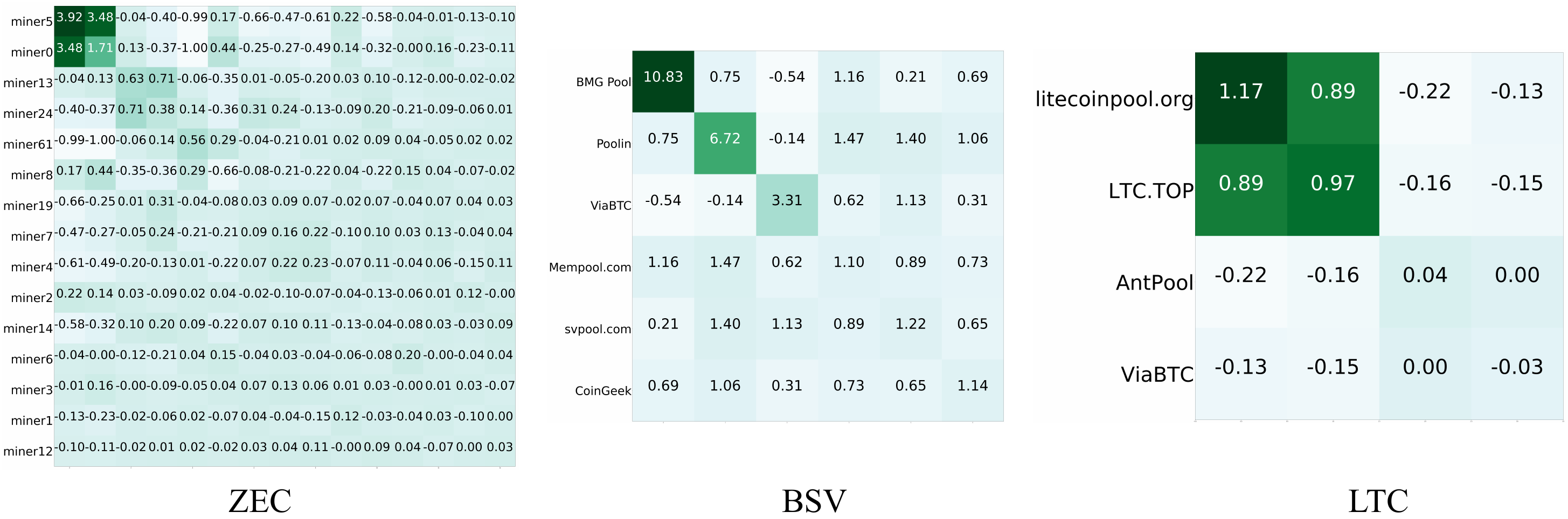}
\caption{These are the pairs matrices, which are used to highlight possible mining cartels in each cryptocurrency.}
\label{fig:pair}
\end{figure*}

\textbf{Results}
Figure~\ref{fig:pair} shows the pairs matrix only for cryptocurrencies that are noteworthy in some way.
In ZEC, we see that miner0 and miner5 are likely part of the same cartel.
Additionally, all BSV miners are getting some previous block advantage, but no explicit cartels 
    have formed.
In LTC, there seems to be another likely cartel between LTC.TOP and litecoinpool.org, but this is 
    not as extreme as the one in ZEC.

Miner centralization, while a problem, is possible to detect and base decisions on, such as 
    users selling off tokens if a coin is too centralized.
However, miner cartels are more insidious as they give the illusion of decentralization while 
    hiding a system that is much more centralized in nature.
Our techniques make it easy to detect cartels so that it is possible to detect and base decisions
    on.

\begin{table*}

  \begin{tabular}{|c|cccccc|}
      \hline
      & Ethermine&SparkPool~&F2Pool\_2~&Nanopool&~\tabincell{c}{MiningPool\\Hub\_1}~& ~\tabincell{c}{BitClub\\Pool}~\\
     estimated hash power& 0.3202 & 0.1867 & 0.1503 & 0.1191 & 0.1015 & 0.0332   \\ 
     corrected estimate &0.3172&0.1866&0.1505&0.1200&0.1022&0.0334\\ 
     estimated hash power using uncles & 0.3108 & 0.1825 & 0.1505 & 0.1332 & 0.1007 & 0.0348 \\
      \hline
      &DwarfPool\_1&bw&miner0&miner10&miner2&\\
      estimated hash power& 0.0220&0.0181&0.0168 & 0.0166 & 0.0148&\\
      corrected estimate&0.0223&0.0184&0.0169&0.0169&0.0150&\\
      estimated hash power using uncles& 0.0224 & 0.0173  &0.0168 & 0.0160 & 0.0145&\\
      \hline
\end{tabular}
  \caption{To validate our debiasing technique to estimate the real mining power, we look at the estimated mining power
    distribution on Ethereum by including uncle blocks. The rows in this table contain, respectively, the identity of
    the miner, the miner's estimated hashpower using just blocks on the blockchain, the de-biased hashpower using our
    technique, and the estimated hashpower using both blocks and uncles.}
  \label{tab:real-mining-power-uncle}
\end{table*}

\subsection{Active Chain Switching}
\label{sec:btctop}

In Nakamoto consensus, miners are rewarded every time they mine a block in the cryptocurrency they are mining.
To remain profitable, miners use specialized hardware that is optimized for the proof of work function for
    their target cryptocurrency.
For example, in our study, a Bitcoin (BTC) miner can easily switch to Bitcoin Cash (BCH) or Bitcion SV (BSV)
    as they all share the same proof of work function.
Thus, miners are incentivized to switch between cryptocurrencies depending on the prices at the time.

Our analysis assumes that the network hashrate is relatively consistent across our measurement period, so
    if miners frequently actively chain switch, that would violate our assumption.
In particular, we assume that taking the sample of blocks on the chain immediately succeeding a miner's blocks
    would give us a representative subsample of all blocks on the chain.
When a miner is actively chain switching, they are more likely to mine after themselves since the fact that a
    block of theirs appears on the chain implies that they are currently mining on this cryptocurrency.
This makes it more likely that they mine the next block, which will show up in our dataset as an exceedingly
    large previous block advantage.

\begin{figure*}
\includegraphics[width=\linewidth]{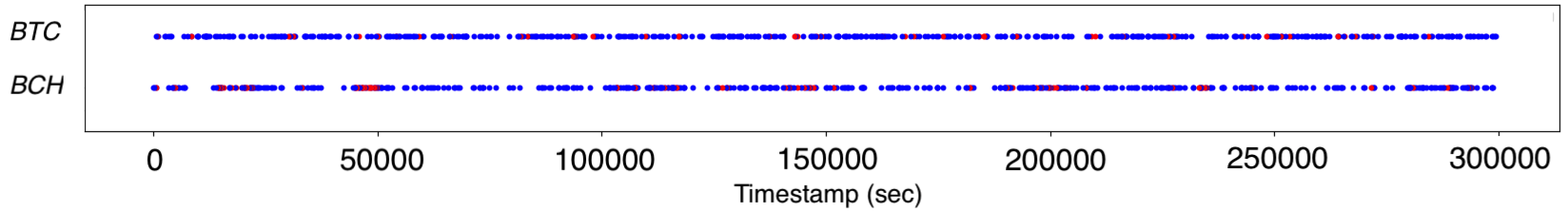}
\caption{We see that \texttt{BTC.TOP} is likely chain switching between BTC and BCH by looking at the timestamps of
when \texttt{BTC.TOP} blocks were mined in BTC and BCH.}
\label{fig:timeline_bch_btc.top}
\end{figure*}

A potential example of such a miner is BTC.TOP in BCH, who may be active chain switching between BTC and BCH.
If we look at BTC.TOP's blocks in Figure~\ref{fig:timeline_bch_btc.top}, we see that few blocks were mined on
    BTC when BTC.TOP was mining BCH and vice versa.
However, this does not fully explain BTC.TOP's large gaps when we look at its mining behavior on BCH as we do
    not see a large previous block advantage for BTC.TOP in BTC.
In general, strategic mining behavior that affects when their blocks appear will be detrimental to our 
    analysis and is a limitation of our techniques.
However, we see that this kind of behavior is not very common in the cryptocurrencies we have studied.

\section{Figures and Tables}
\label{app:fig}

In this section, we include more detailed figures and tables to support the claims made in the main body of the paper.

\begin{figure*}
\centering
\includegraphics[width=\linewidth]{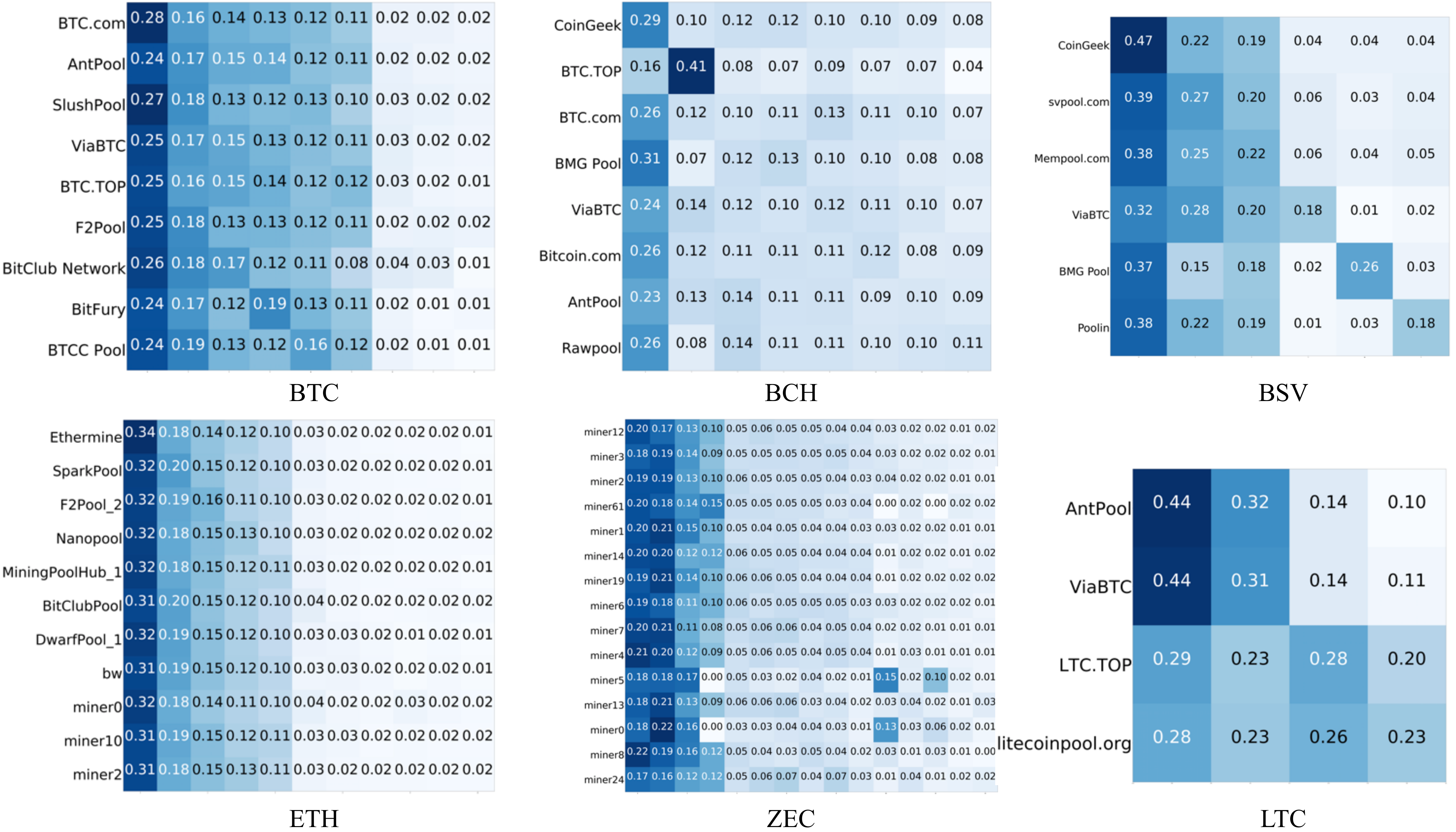}
\caption{These are the succession matrices.}
\label{fig:freq_p}
\end{figure*}
\begin{figure*}
\centering
\includegraphics[width=\linewidth]{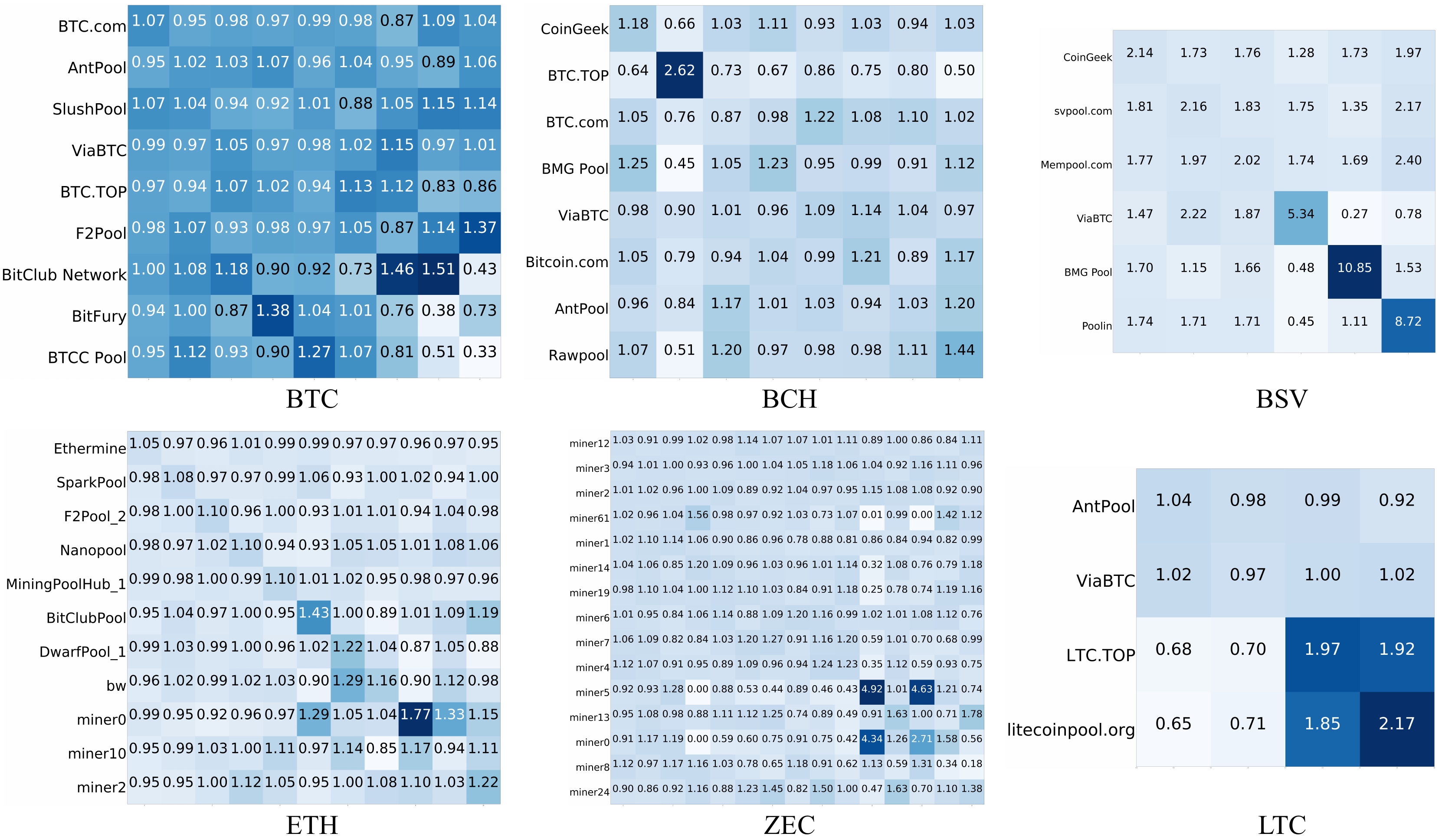}
\caption{We present normalized succession matrices for each cryptocurrency to highlight exactly how much of an advantage or
    disadvantage a miner gets based on the identity of the previous block's miner.}
\label{fig:freq_d}
\end{figure*}

\end{document}